\documentclass[fleqn,usenatbib,useAMS]{mnras}

\usepackage{newtxtext,newtxmath}

\usepackage[T1]{fontenc}
\usepackage{ae,aecompl}

\usepackage[dvipdfmx]{graphicx}	
\usepackage{amsmath}	
\usepackage{amssymb}    
\usepackage{bm}		    
\usepackage{pdflscape}	
\usepackage{color}

\title[Detailed study of detection method for ultra-low-frequency GWs]{Detailed study of detection method for ultra-low-frequency gravitational waves with pulsar spin-down rate statistics}

\author[S.Hisano et al.]{
Shinnosuke Hisano,$^{1}$\thanks{E-mail: 184d8062@st.kumamoto-u.ac.jp}
Naoyuki Yonemaru,$^{1, 2}$
Hiroki Kumamoto,$^{1, 2}$ \newauthor
and Keitaro Takahashi$^{1}$
\\
$^{1}$Kumamoto University, Graduate School of Science and Technology, Kumamoto, 860-8555, Japan\\
$^{2}$CSIRO Astronomy and Space Science, PO Box 76, Epping NSW 1710, Australia\\
}

\date{Accepted XXX. Received YYY; in original form ZZZ}

\pubyear{2019}

\begin{document}
\label{firstpage}
\pagerange{\pageref{firstpage}--\pageref{lastpage}}
\maketitle

\begin{abstract}
A new detection method for gravitational waves (GWs) with ultra-low frequencies ($f_{\rm GW} \lesssim 10^{-10}~{\rm Hz}$), which is much lower than the range of pulsar timing arrays (PTAs), was proposed in Yonemaru et al. (2016). This method utilizes the statistical properties of spin-down rates of milli-second pulsars (MSPs) and the sensitivity was evaluated in Yonemaru et al. (2018). There, some simplifying assumptions, such as neglect of the "pulsar term" and spatially uniform distribution of MSPs, were adopted and the sensitivity on the time derivative of GW amplitude was estimated to be $10^{-19}~{\rm s}^{-1}$ independent of the direction, polarization and frequency of GWs. In this paper, extending the previous analysis, realistic simulations are performed to evaluate the sensitivity more reasonably. We adopt a model of 3-dimensional pulsar distribution in our Galaxy and take the pulsar term into account. As a result, we obtain expected sensitivity as a function of the direction, polarization and frequency of GWs. The dependence on GW frequency is particularly significant and the sensitivity becomes worse by a few orders for $< 10^{-12}~{\rm Hz}$ compared to the previous estimates.
\end{abstract}




\section{Introduction}

Pulsars emit pulses with a very stable period so that they can be used as precise clocks and the arrival time of pulses can be predicted. In the presence of gravitational waves (GWs), the pulse propagation path changes and the arrival time of pulses deviate from the prediction. Utilising this phenomenon, low frequency GWs with frequencies $10^{-9} - 10^{-6}~{\rm Hz}$ can be detected and this method is called pulsar timing array (PTA) \citep{Foster}.

Currently, there are three PTAs in operation; the Parkes PTA in Australia \citep{Manchester2012}, the European PTA \citep{Kramer}, and NANOGrav in the United States \citep{2013CQGra..30v4008M}. Further, in the 2020s, the Square Kilometre Array (SKA) will start running \citep{Kramer2015} and 1,400 and 3,000 milli-second pulsars (MSPs) are expected to be discovered by the SKA1 and SKA2 surveys, respectively \citep{Keane}. So far, PTAs have put constraints on the gravitational wave background (GWB) \citep{Shannon,2015MNRAS.453.2576L,2018ApJ...859...47A} and GWs from individual supermassive black hole (SMBH) binaries \citep{2014MNRAS.444.3709Z,2016MNRAS.455.1665B,2018arXiv181211585A}. Adding more pulsars regularly to PTAs will continually improve detection probability \citep{2016ApJ...819L...6T} and the GWB from SMBH binaries will be strongly expected to be detected with these observations in the future.

The frequency range of GWs probed by PTAs is determined by the observational time span (O(10) years) and cadence (O(1) week). In this frequency range, targets are SMBH binaries and their gravitational wave background. Concerning SMBH binaries, the frequency range of PTAs corresponds to the sub-pc orbital radii (e.g. $6 \times 10^{-3}$ pc for an equal mass binary of $10^{8}$ M$_{\odot}$ leads to GWs of $10^{-8}$ Hz). A SMBH binary with milli-pc orbital radii corresponds to the late stage of the evolution. On the other hand, in the early stage, the orbit of a SMBH binary shrinks with the extraction of angular momentum by scattering of stars and the friction of surrounding gas. However, when the orbital radius is a few pc, the transfer of angular momentum by stars and gas is no longer effective. Thus, it takes a longer time than the Hubble time for two SMBHs to merge only by GW emission \citep{Merritt,Lodarto}. This is called "the final parsec problem" and it is important to observe pc-scale binaries to solve this problem. However, GWs with a ultra-low frequency $\ll 10^{-9}$ Hz, which should be emitted by the early stage of SMBH binaries, cannot be detected with the conventional method of PTAs. This is because the pulsar spin-down rate and ultra-low frequency GWs have the same time dependence in time of arrival and such GWs are absorbed by the parameter fitting of the pulsar spin-down rates.

In our previous work \citep{Yonemaru2016}, we proposed a new detection method for such ultra-low-frequency GWs. The method utilizes the dependence of the GW effect on the sky position of MSPs and GWs can be probed through the statistical properties of observed spin-down rates of multiple MSPs. Then, in \cite{Yonemaru2018}, we evaluated the sensitivity of this method in a simple manner and it was concluded that GWs with the time derivative of amplitude of order $10^{-19}~{\rm s}^{-1}$ could be probed. There, it was assumed that MSPs are located uniformly in the sky and the "pulsar term" was neglected. Under these assumptions, the sensitivity does not depend on the direction, polarization and frequency of GWs.

In fact, pulsar distribution is far from uniform in the sky and should be concentrated on the Galactic plane where most of the Galactic mass exists. Also, the pulsar term is necessary to evaluate the sensitivity more precisely, especially when the wavelength of GWs is comparable to or longer than the typical distance to pulsars as we show below. Thus, in this paper, we consider a realistic model of 3-dimensional pulsar distribution in our Galaxy and take the pulsar term into account in order to obtain better estimates of the sensitivity, extending the analysis in \cite{Yonemaru2018}.

This paper is organized as follows. In section \ref{Detection principle}, we describe the detection principle of the new method for ultra-low-frequency GWs. In section \ref{Simulation}, we show the setup of our simulations to assess the sensitivity. The results and interpretation are presented in section \ref{Results}. In section \ref{Summary}, we give a summary.

\section{Detection principle}
\label{Detection principle}

In the presence of GWs, actual time of arrival of pulses from a pulsar is deviated from the expectation without GWs. This difference is called timing residual and given by \cite{Detweiler},
\begin{equation}
  r_{\rm{GW}}(t) = \sum_{A=+.\times} F^A (\hat{\bf \Omega},\hat{\mbox{\boldmath $p$}}) \int^t \Delta h_A(t^{\prime},\hat{\bf \Omega},{\theta}) dt^{\prime},
\end{equation}
where $\hat{\bf \Omega}$, $\hat{\mbox{\boldmath $p$}}$ are the directions of the pulsar and GW propagation, respectively, and $\theta$ is the GW polarization angle. Here, $F^A (\hat{\bf \Omega},\hat{\mbox{\boldmath $p$}})$ is the antenna beam pattern given by \cite{Anholm}
\begin{equation}
F^A (\hat{\bf \Omega},\hat{\mbox{\boldmath $p$}}) =  \frac{1}{2} \frac{\hat{\mbox{\boldmath $p$}^i}\hat{\mbox{\boldmath $p$}^j}}{1+\hat{\bf \Omega}\cdot\hat{\mbox{\boldmath $p$}}}\mbox{\boldmath $e$}_{ij}^A(\hat{\bf \Omega}),
\end{equation}
where $\mbox{\boldmath $e$}_{ij}^A(\hat{\bf \Omega})(A=+,\times)$ is the GW polarization tensor,
\begin{eqnarray}
\mbox{\boldmath $e$}_{ij}^+(\hat{\bf \Omega}) &=& \hat{\mbox{\boldmath $m$}_i}\hat{\mbox{\boldmath $m$}_j}-\hat{\mbox{\boldmath $n$}_i}\hat{\mbox{\boldmath $n$}_j} \\
\mbox{\boldmath $e$}_{ij}^\times(\hat{\bf \Omega}) &=& \hat{\mbox{\boldmath $m$}_i}\hat{\mbox{\boldmath $n$}_j}+\hat{\mbox{\boldmath $n$}_i}\hat{\mbox{\boldmath $m$}_j}
\end{eqnarray}
where $\hat{\mbox{\boldmath $m$}}$ and $\hat{\mbox{\boldmath $n$}}$ are the polarization basis vectors given by
\begin{eqnarray}
\hat{\mbox{\boldmath $m$}} &=& (\sin{\phi},-\cos{\phi},0) \\
\hat{\mbox{\boldmath $n$}} &=& (\sin{\psi}\cos{\phi},-\cos{\psi}\sin{\phi},-\sin{\theta}).
\end{eqnarray}
Here, $\psi$ and $\phi$ are the galactic longitude and galactic latitude, respectively, and $\Delta h_A(t^{\prime},\hat{\bf \Omega},{\theta})$ is the difference of the metric perturbation between the Earth and pulsar and given by
\begin{equation}
\Delta h_A(t,\hat{\bf \Omega},{\theta}) = h_A(t,\hat{\bf \Omega},{\psi}) - h_A(t_p,\hat{\bf \Omega},{\theta}),
\label{eq:h}
\end{equation}
where $t_p = t - \tau = t - L/c(1 + \hat{\bf \Omega} \cdot \hat{\mbox{\boldmath $p$}})$, $\tau$ is the pulse propagation time to the Earth from the pulsar, and $L$ is the distance to the pulsar. In Eq. (\ref{eq:h}) the first term is "the Earth term" and the second term is "the pulsar term". In our previous work \citep{Yonemaru2018}, we neglected the pulsar term because it contributes as random noise whose average is zero when the GW wavelength is much shorter than the typical pulsar distance (GW frequency is much larger than $10^{-13}$Hz). In this study we take the pulsar term into account in order to estimate the sensitivity of this method more precisely.

The Earth and pulsar terms have almost the same frequency when the GW source is a SMBH binary with such a low frequency \citep{Yonemaru2018}, and we can rewrite Eq. (\ref{eq:h}) as    
\begin{equation}
\Delta h_A(t,\hat{\bf \Omega},\theta) = \dot{h}_A t \,(1-e^{2i\pi f_{\rm{GW}}\tau}).
\label{eq:delta-h}
\end{equation}
Then timing residual is given by,
\begin{equation}
r_{\rm{GW}}(t) = \frac{1}{2} \sum_{A=+.\times} F^A (\hat{\bf \Omega},\hat{\mbox{\boldmath $p$}})(1-e^{2i\pi f_{\rm{GW}}\tau})\dot{h}_A t^2 
\end{equation}
However, as stated above, such effect is absorbed by shifting the pulsar spin-down rate. In other words, the spin-down rate is biased by ultra-low frequency GWs. Denoting the bias factor as $\alpha(\hat{\bf \Omega},\hat{\mbox{\boldmath $p$}},\theta)$, the observed spin-down rate $\dot{p}$ is given by
\begin{equation}
\frac{\dot{p}}{p} = \frac{\dot{p_0}}{p} + \alpha(\hat{\bf \Omega},\hat{\mbox{\boldmath $p$}},\theta),
\end{equation}
where $p$ is the pulse period, $\dot{p}_0$ is the intrinsic spin-down rates. Here the bias factor is given by
\begin{equation}
\alpha(\hat{\bf \Omega},\hat{\mbox{\boldmath $p$}},\theta) = \sum_{A=+.\times} F^A (\hat{\bf \Omega},\hat{\mbox{\boldmath $p$}}) (1-e^{2i \pi f_{\rm{GW}} \tau}) \dot{h}_A,
\label{eq:alpha}
\end{equation}
where $\dot{h}_+$ and $\dot{h}_\times$ are given by
\begin{eqnarray}
\dot{h}_+ &=& \dot{h}\cos{2\theta} \\
\dot{h}_{\times} &=& \dot{h}\sin{2\theta}.
\end{eqnarray}

\begin{figure}
\centering
\includegraphics[width=8cm,bb=0 0 370 300]{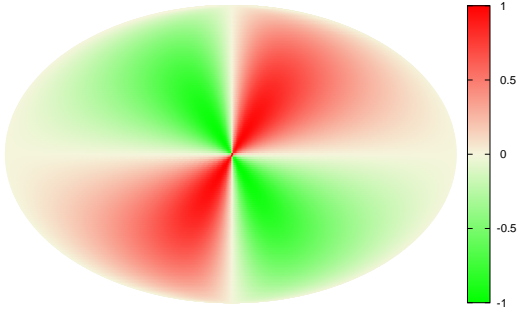}
\caption{Antenna beam pattern $F^{\times} (\hat{\bf \Omega},\hat{\mbox{\boldmath $p$}})$ in the sky. It is assumed that the source exists in the direction of the Galactic Center (GC).}
\label{fig:antenna}
\end{figure}

In principle, we cannot extract the GW effect from the observed spin-down rate of a single pulsar and, thus, such ultra-low-frequency GWs cannot be detected by the conventional PTA method. However, because the bias depends on the relative position of the GW source and pulsar, such low-frequency GWs could be probed through the statistical properties of spin-down rates of multiple pulsars in the sky. Actually, if there is only one GW source, the spatial pattern of the bias factor $\alpha(\hat{\bf \Omega},\hat{\mbox{\boldmath $p$}},\theta)$ is quadrupole as shown in Fig. \ref{fig:antenna}.

\begin{figure}
\centering
\includegraphics[width=\linewidth]{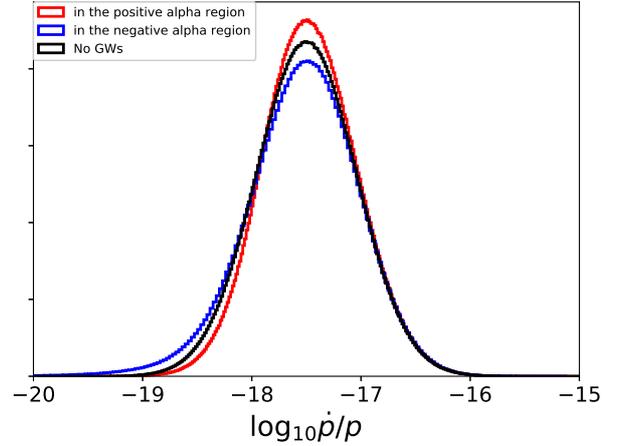}
\caption{Schematic view of the histogram of $\log_{10}\dot{p}/{p}$ with $\dot{h}_{\times}=10^{-17}\rm{s}^{-1}$. The black line is the assumed Gaussian distribution with the mean and variance which are the same as the values mentioned in section \ref{Detection principle}. The red and blue lines show the distribution in the areas with positive and negative $\alpha(\hat{\bf \Omega},\hat{\mbox{\boldmath $p$}},\theta)$, respectively.}
\label{pdotp}
\end{figure}

In our previous work \citep{Yonemaru2018}, we proposed to utilize the skewness of spin-down rate distribution of two sky areas with positive and negative values of $\alpha(\hat{\bf \Omega},\hat{\mbox{\boldmath $p$}},\theta)$. Let us review the method briefly here. First, assuming the direction of GW source and polariztion, the sky is separated into two areas according to the sign of $\alpha(\hat{\bf \Omega},\hat{\mbox{\boldmath $p$}},\theta)$. Then we obtain the histogram of $\log_{10}\dot{p}/{p}$ of pulsars in each area. In the area with positive (negative) $\alpha(\hat{\bf \Omega},\hat{\mbox{\boldmath $p$}},\theta)$, $\dot{p}/{p}$ of the pulsars have a positive (negative) bias and the distribution tends to be positively (negatively) skewed because the left-hand-side tail of the distribution is shortened (extended) as shown in Fig. \ref{pdotp}. Finally, the skewness is calculated for each area. Here, the skewness is defined as 
\begin{equation}
S = \frac{1}{\sigma^{3} N} \sum_{i}^{N}\biggl(\log_{10}\biggl(\frac{\dot{p}}{p}\biggr)_i-\mu \biggr)^{3},
\end{equation}
where $N$ is the number of MSPs, and $\mu$ and $\sigma$ are the mean and standard deviation of the $\log_{10}\dot{p}/{p}$ distribution, respectively,
\begin{eqnarray}
&& \mu = \frac{1}{N} \sum_{i}^{N} \log_{10}\biggl(\frac{\dot{p}}{p}\biggr)_i, \\
&& \sigma^{2} = \frac{1}{N} \sum_{i}^{N} \biggl(\log_{10}\biggl(\frac{\dot{p}}{p}\biggr)_i-\mu \biggr)^{2}.
\end{eqnarray}
It was shown that the difference of the skewness could be a good measure of amplitude of ultra-low-frequency GWs which induce the skewness difference larger than the statistical error.

For currently known 149 MSPs with $p < 30~{\rm ms}$ (the ATNF pulsar catalogue: \citep{Manchester}), the mean and variance of $\log_{10}\dot{p}/{p}$ are -17.5 and 0.21, respectively. Here, MSPs in globular clusters are excluded because they are biased significantly by the gravitational potential and complicated dynamics inside the cluster. Further, the observed spin-down rates are affected by Shklovskii effect \citep{Shkloviskii} and acceleration along the sight by the Galactic differential rotation \citep{Damour,Rong} other than GWs. However, such effects do not have spatial correlations in the sky and do not affect our analysis below. 

\section{Simulation}
\label{Simulation}

In this section, we describe the simulations for estimating the sensitivity of our method more reasonably than our previous work. To do so, we consider the 3-dimensional spatial distribution of MSPs in Galaxy, while MSPs were assumed to be distributed uniformly in the sky and pulsar term, which involves the distance to the MSPs, was neglected in our previous work. Concerning the value of $\log_{10}\dot{p}/{p}$, we continue to assume Gaussian distribution with the mean and variance of -17.5 and 0.21, respectively.

\begin{figure}
\centering
\includegraphics[width=8cm,bb=0 0 370 300]{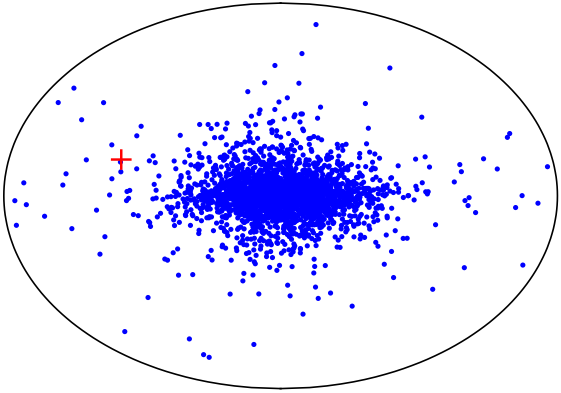}
\caption{Example distribution of 3,000 MSPs in the sky seen from the Earth. The symbol '+' represents the position of M87.}
\label{distribution}
\end{figure}

\begin{figure}
\centering
\includegraphics[width=\linewidth]{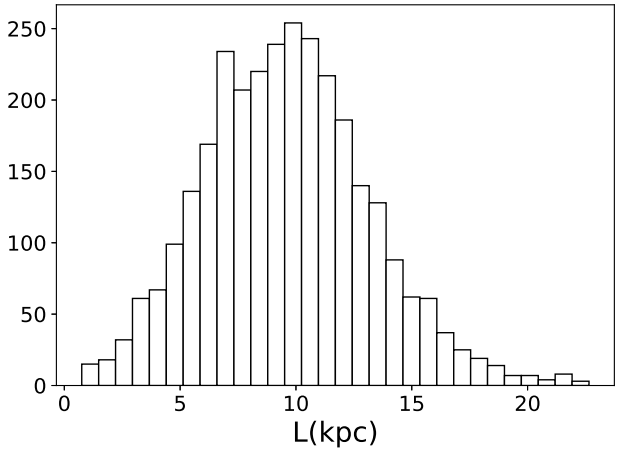}
\caption{Example distribution of the pulsar distance from the Earth.}
\label{L}
\end{figure}

We simulate the spatial distribution of MSPs using a model developed in \cite{Lorimer,Paul}. This model is based on a pulsar population synthesis code that accounts for isolated and binary pulsar evolution, taking Galactic spatial evolution and pulsar survey selection effects into account. There, MSPs are considered to be formed through mass and angular momentum transfer onto a neutron star from an evolving giant companion and the resultant companion is likely a low-mass white dwarf (WD). The radial density profile is given by
\begin{equation}
\rho(r) = 4.1~{\rm kpc}^{-2} \biggl(\frac{r}{R_{{\odot}}}\biggr)^{1.9} \exp \biggl(-5.0\biggl[\frac{r-R_{\odot}}{R_{\odot}}\biggr]\biggr),
\end{equation}
where $r~{\rm [kpc]}$ is the distance from the GC, ${R_{\odot}} = 8.5~{\rm kpc}$ is the Sun-GC distance and $\rho(r)$ is the number of MSPs per unit area on Galactic plane, while the azimuth angle is given randomly. On the other hand, the distribution perpendicular to the Galactic plane is given by
\begin{equation}
N(z) = 0.75~{\rm kpc}^{-1} \exp(-|z|/0.83).
\end{equation}
where $z~{\rm [kpc]}$ is the height from the Galactic plane. We put 3,000 MSPs according to these distribution function and obtain their sky positions and distances from the Earth. Figs \ref{distribution} and \ref{L} show an example distribution in the sky and the histogram of the distance, respectively. We find that MSPs are concentrated on the Galactic plane, which is very different from the assumption of our previous work. 

Once the sky positions and distances of MSPs are given, the bias factor $\alpha(\hat{\bf \Omega},\hat{\mbox{\boldmath $p$}},\theta)$ can be allocated to each MSP, assuming the direction, polarization, frequency and amplitude of GW. In order to evaluate the detectability, following \cite{Yonemaru2018}, we estimate the probability distribution of skewness difference of the two areas with positive and negative value of $\alpha(\hat{\bf \Omega},\hat{\mbox{\boldmath $p$}},\theta)$ through multiple realizations. It should be noted that, in a practical observation, we need to seek for the direction and polarization which give the largest skewness difference.

In our simulations, we assume the intrinsic $\log_{10}\dot{p}/{p}$ distribution is Gaussian. In fact, the Gaussianity of the current samples was tested by the Jarque-Bera test and, with the skewness $S=1.2$ and kurtosis $K=5.9$, the distribution is inconsistent with Gaussianity with more than 99.9$\%$ significance \citep{Yonemaru2018}. However, it was shown that the intrinsic skewness is largely canceled when we take the difference between the two areas and does not affect the results so much.

\section{Results}
\label{Results}

\begin{figure*}
\centering
\includegraphics[width=150mm]{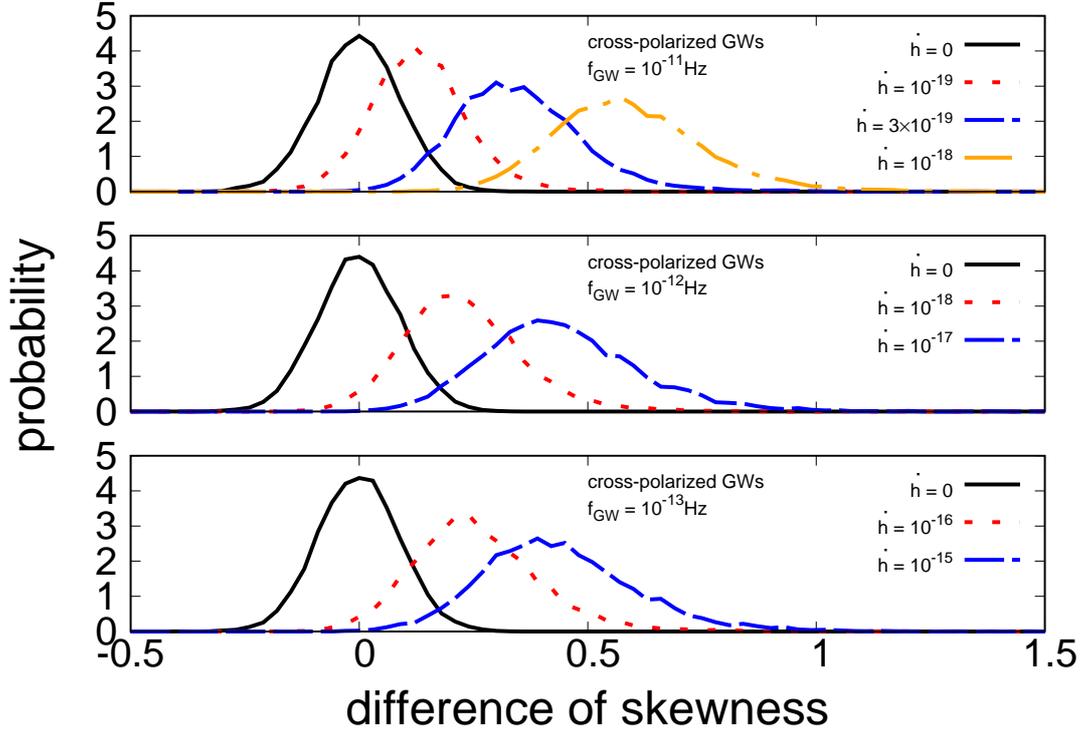}
\caption{Probability distribution of the skewness difference of the $\log_{10}\dot{p}/{p}$ distributions with cross-polarized GWs $\theta = 45^{\circ}$ for $f_{\rm{GW}} = 10^{-11}$ (top), $10^{-12}$ (middle) and $10^{-13}~{\rm Hz}$ (bottom).}
\label{cross}
\end{figure*}

\begin{figure*}
\centering
\includegraphics[width=150mm]{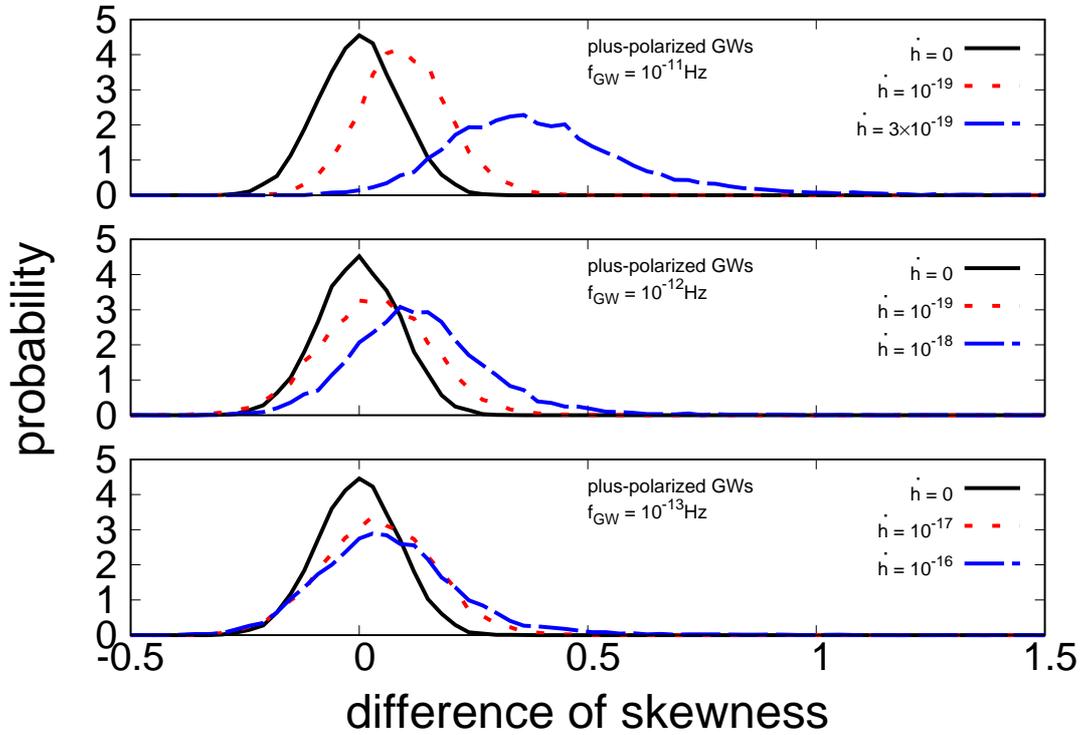}
\caption{Same as Fig. \ref{cross} for plus polarization $\theta = 0^{\circ}$.}
\label{plus}
\end{figure*}

When the pulsar term is neglected and MSPs are distributed uniformly in the sky, as in the case of \cite{Yonemaru2018}, the sensitivity, which is determined by the probability distribution of skewness difference, is independent of the frequency, direction and polarization of GWs. However, we need to consider these dependence in our case. We primarily assume the GW is from the direction of the GC, while we consider the direction of M87 as well. In the former case, the numbers of MSPs which has positive and negative values of $\alpha(\hat{\bf \Omega},\hat{\mbox{\boldmath $p$}},\theta)$ drastically change depending on the GW polarization. As we see in Figs. \ref{fig:antenna} and \ref{distribution}, the numbers are almost the same for the case of cross polarization, while the number ratio of positive and negative areas is about 1:5 or 5:1 for the case of plus polarization. The statistical fluctuation is larger and, thus, the sensitivity depends strongly on the polarization angle of the GW. Below, we show these extreme cases.

First, let us show the case with cross polarization ($\theta = 45^{\circ}$). Fig. \ref{cross} shows the probability distribution of the skewness difference with and without GWs for $f_{\rm GW} = 10^{-11}~{\rm Hz}$, $10^{-12}~{\rm Hz}$ and $10^{-13}~{\rm Hz}$. The average value of the probability distribution is zero without GWs and increases as the value of $\dot{h}_{\times}$ increases. The probability distribution has statistical fluctuations of about 0.2 for a fixed value of $\dot{h}_{\times}$ and it tends to increase slightly for larger values of $\dot{h}_{\times}$.

Focusing on the GW-frequency dependence, we notice that the skewness difference reduces very rapidly for decreasing the frequency. The top panel with $f_{\rm GW} = 10^{-11}~{\rm Hz}$ is quantitatively similar to the result of our previous work. On the other hand, for $f_{\rm GW} = 10^{-13}~{\rm Hz}$, the probability distribution in case of $\dot{h} \leq 10^{-17}$ is almost the same as the case without GWs, which implies that such GWs cannot detected by this method. We will give an interpretation to the frequency dependence in the next section.

Similarly, Fig. \ref{plus} shows the probability distribution of the skewness difference for plus-polarized GWs ($\theta = 0^{\circ}$). Because the number of MSPs of either positive or negative area is very small, the skewness difference tends to be small compared with the case of cross polarization. The difference between the two polarization cases is significant especially for low GW frequencies.

\begin{figure}
\centering
\includegraphics[width=\linewidth]{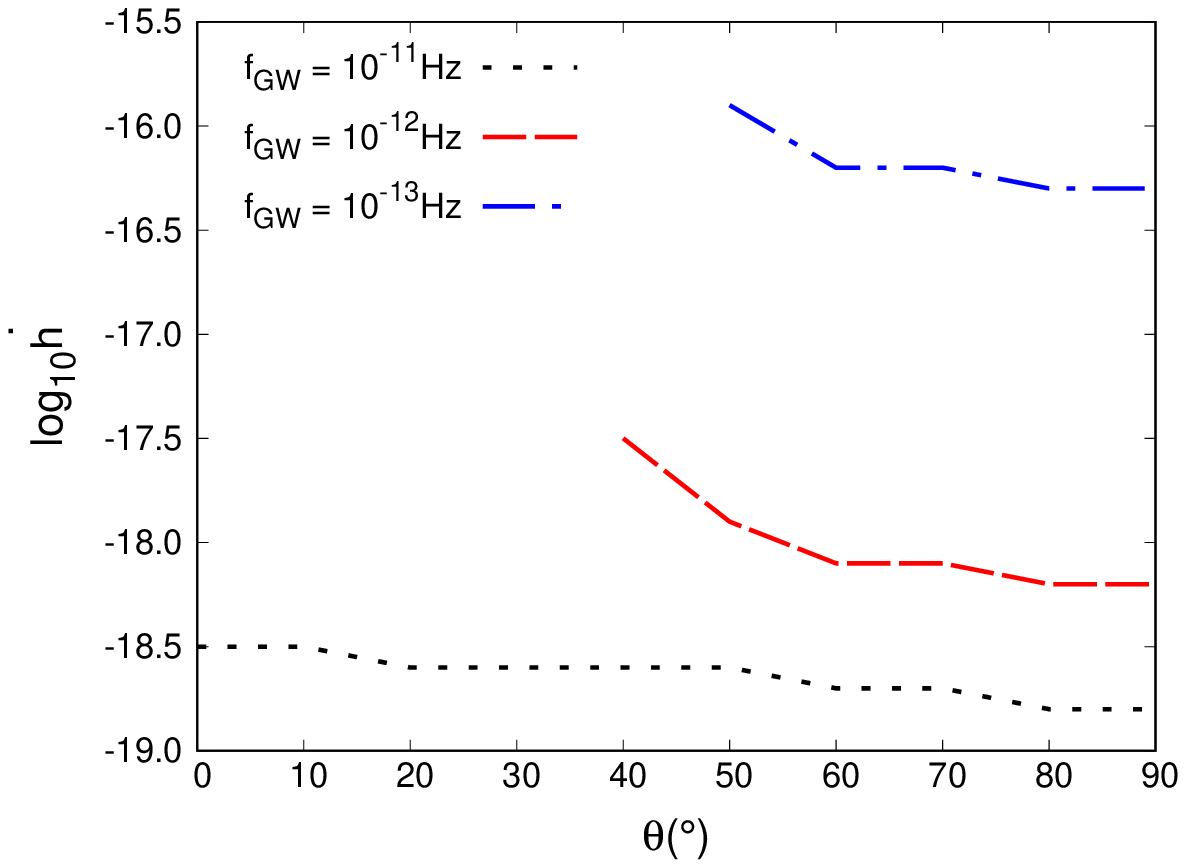}
\caption{Sensitivity as a function of polarization angle for $f_{\rm{GW}} = 10^{-11}$, $10^{-12}$ and $10^{-13}~{\rm Hz}$. Here, $\theta = 0^{\circ}$ and $90^{\circ}$ correspond to plus polarization, while $\theta = 45^{\circ}$ corresponds to cross polarization.}
\label{sensitivity-GC}
\end{figure}

We define the sensitivity of GWs as follows. First, for a fixed value of GW frequency and polarization, we evaluate the value of skewness difference with $\dot{h}=0$, $S_{\rm c}$ as,
\begin{equation}
\int_{S_{\rm c}}^{\infty} P(S,\dot{h}=0)~dS = 0.05
\end{equation}
where $P(S,\dot{h})$ is the probability distribution of skewness difference. Then, GWs with $\dot{h}$ are considered to be detectable if
\begin{equation}
\int_{S_{\rm c}}^{\infty} P(S,\dot{h})~dS \geq 0.9
\end{equation}
is satisfied. The sensitivity is defined as the least value of $\dot{h}$ which satisfies the above equation.

Fig. \ref{sensitivity-GC} shows the sensitivity defined above as a function of polarization angle for $f_{\rm{GW}} = 10^{-11}$, $10^{-12}$ and $10^{-13}~{\rm Hz}$. First, let us see the case with $f_{\rm{GW}} = 10^{-11}~{\rm Hz}$. The sensitivity is slightly better for larger polarization angle. Especially, although both $\theta = 0^{\circ}$ and $90^{\circ}$ correspond to plus polarization, the sensitivity is better for the latter. In the case with  $\theta = 0^{\circ} (90^{\circ})$, most of the Galactic plane has positive (negative) value of the bias factor $\alpha$. The effect of GWs on skewness is greater for the negative-$\alpha$ area because skewness is the third moment around the mean and the absolute value of skewness is larger with a larger number of pulsars which have the value of $\log_{10}\dot{p}/{p}$ significantly deviated from the mean. Thus, the sensitivity is better for the case with $\theta = 90^{\circ}$ where most of pulsars are located in negative-$\alpha$ area. It should be noted that there is no such asymmetry between two cross polarizations ($\theta = 45^{\circ}$ and $135^{\circ}$). 

\begin{figure}
\centering
\includegraphics[width=\linewidth]{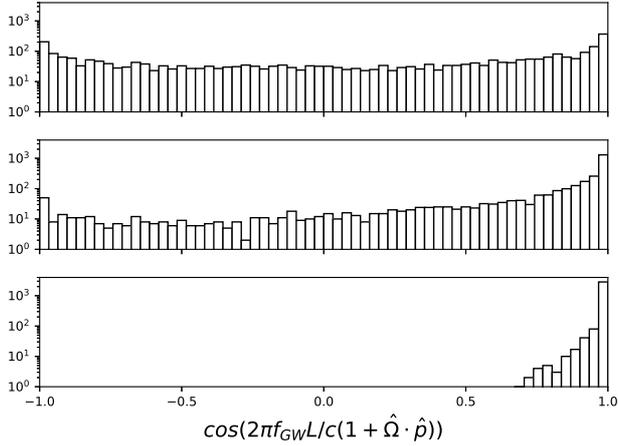}
\caption{Histogram of the pulsar term for $f_{\rm{GW}}=10^{-11}$Hz (top), $f_{\rm{GW}}=10^{-12}$Hz (middle) and $f_{\rm{GW}}=10^{-13}$Hz (bottom).}
\label{cos_-11}
\end{figure}

As to the frequency dependence, it is seen that the sensitivity becomes worse rapidly for lower frequencies. To understand this, let us see more details of behavior of the probability distribution of the skewness difference at low frequencies $f_{\rm{GW}} \leq 10^{-12}$. We focus on the factor $[1 - \exp{(2i\pi f_{\rm GW} \tau)}]$ in Eq.(\ref{eq:alpha}), where $\tau = (1 + \hat{\bf \Omega} \cdot \hat{\mbox{\boldmath $p$}})L/c$, and the first and second terms come from the Earth and pulsar terms, respectively. It is seen that the pulsar term can cancel the Earth term depending on the phase, which is determined by the relative direction of the pulsar and GW source, GW frequency and the pulsar distance.  Especially, when the directions of the pulsar and GW source are very close to each other, $\tau \sim 0$ and the pulsar term cancels out the Earth term. This effect is known as the "surfing effect" \citep{Braginsky} and is significant for $\tau \ll 1$, that is,
\begin{equation}
1 - \cos{\beta} \ll \frac{c}{\pi f_{\rm GW} L}
\end{equation}
where $\beta$ is the angle between the pulsar and GW source. Therefore, the sky area where pulsars have $\tau \sim 0$ is larger for lower frequencies. This is why the sensitivity is worse for lower frequencies due the smaller number of pulsars which have non-negligible value of $\tau$. This effect is even more serious in the case that the GW source is in the direction of GC, around which most of pulsars are located. In terms of the bias factor, when $\tau \ll 1$, we have
\begin{equation}
1 - e^{(2i\pi f_{\rm GW} \tau)}
\simeq  2\pi^{2}f_{\rm{GW}}^{2}L^{2}/c^{2} (1+\hat{\bf \Omega}\cdot\hat{\mbox{\boldmath $p$}})^{2},
\end{equation}
so that  
\begin{equation}
\alpha(\hat{\bf \Omega},\hat{\mbox{\boldmath $p$}},\theta) = \sum_{A=+.\times}\hat{p^i}\hat{p^j}\,e_{ij}^A(1+\hat{\bf \Omega}\cdot\hat{\mbox{\boldmath $p$}}) \pi^2 f_{\rm{GW}}^2 (L^2 / c^2) \dot{h}_A.
\end{equation}
Fig. \ref{cos_-11} shows examples of the histogram of the real part of the pulsar term for $f_{\rm{GW}} = 10^{-11}, 10^{-12}$ and $10^{-13}~{\rm Hz}$. In the case with $f_{\rm GW} = 10^{-12}$ and $10^{-13}~{\rm Hz}$, the wavelength of the GW (10kpc and 100kpc, respectively) is larger than the distance of many pulsars (see Fig. \ref{L}) so that the real part of the pulsar term is mostly unity and it cancels the Earth term.

Further, the sensitivity curve is terminated at $\theta = 40^{\circ}$ and $50^{\circ}$ for $f_{\rm GW} = 10^{-12}$ and $10^{-13}~{\rm Hz}$, respectively. This is because, when the value of $\dot{h}$ is too large, the distribution of $\log_{10}\dot{p}/{p}$ is far from Gaussian and skewness is no longer a reasonable quantity to characterize the distribution. In that case, we will need a different indicator of the GW amplitude instead of the skewness difference but this is beyond the scope of the current paper.

\begin{figure}
\centering
\includegraphics[width=\linewidth]{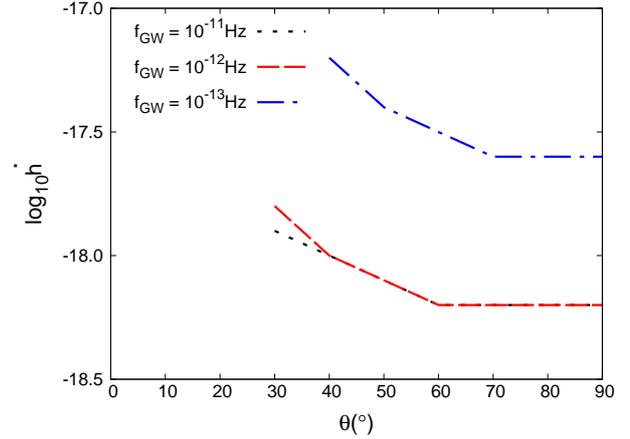}
\caption{Same as Fig. \ref{sensitivity-GC} but in the direction of M87.}
\label{sensitivity-M87}
\end{figure}

Finaly, we consider the case that GWs come from the direction of M87. There is a possibility that another SMBH other than the one associated with the AGN \citep{Batcheldor}. In \cite{Yonemaru2016}, GWs from the SMBH binary was considered and the frequency of GWs is too low to be detected by the conventional PTA. The sensitivity for the case of M87 is shown in Fig. \ref{sensitivity-M87}. For $\theta \lesssim 30^{\circ}$, the current method is insensitive to GWs because most of MSPs are located in positive-$\alpha$ area and the skewness difference is not statistically significant. On the other hand, for $\theta \gtrsim 30^{\circ}$, the sensitivity improves significantly in the case of $f_{\rm GW} = 10^{-13}~{\rm Hz}$, while the change is moderate in the other two cases.

\section{Summary}
\label{Summary}

In this paper, we estimated the detection probability of ultra-low-frequency GWs from a single source. We used statistics of observed spin-down rates of MSPs and GW signal appears as the difference of skewness between the spin-down rate distributions of two MSP groups. While we neglected the pulsar term and assumed MSPs were distributed uniformly in the sky in our previous study \citep{Yonemaru2018}, we used a realistic distribution model of MSPs in our Galaxy and took the pulsar term into account. These new ingredients induced the dependence of the sensitivity on the direction, polarization and frequency of GWs. The frequency dependence is especially significant and the sensitivity is degraded by a few orders for $f_{\rm GW} \leq 10^{-12}~{\rm Hz}$ compared to the previous estimate. However, our method is still unique in probing GWs with ultra-low frequencies $f_{\rm GW} \leq 10^{-10}~{\rm Hz}$, which cannot be accessed by the conventional PTAs.

In this paper, we assumed the source is located in the direction of the GC. When Sgr $\rm{A}^{*}$ is the GW source, we need a slight modification to the formalism because the pulsar term in Eq. (\ref{eq:delta-h}) was derived assuming the GWs are plane waves. It is easy to calculate the phase difference of GWs between Earth and pulsars considering the spherical waves from Sgr $\rm{A}^{*}$. This modification will not affect the results of the current paper.

\section*{Acknowledgements}

NY was financially supported by the Grant-in-Aid from the Overseas Challenge Program for Young Researchers of JSPS. KT is partially supported by JSPS KAKENHI Grant Numbers JP15H05896, JP16H05999 and JP17H01110, and Bilateral Joint Research Projects of JSPS.












\bsp	
\label{lastpage}
\end{document}